\documentclass[letterpaper, 10 pt, conference]{ieeeconf}

\IEEEoverridecommandlockouts
\usepackage[utf8]{inputenc}
\usepackage[T1]{fontenc}
\usepackage[english]{babel}

\usepackage{microtype}                      
\usepackage{xparse}                         

\usepackage{amsmath, amssymb}               
\usepackage{amsthm}
\usepackage{mathtools}                      
\usepackage[casesstyle, subnum]{cases}      
\usepackage[per-mode=fraction]{siunitx}     
\usepackage{bm}                             

\usepackage{enumerate}                      
\usepackage{booktabs}                       
\usepackage{multirow}

\makeatletter
\let\NAT@parse\undefined
\makeatother

\usepackage{csquotes}                       
\usepackage{hyperref}
\usepackage{cite}                           

\usepackage{tikz}
\usepackage{pgfplots}                       
\usepackage{pgfplotstable}                  
\usepackage[caption=false,font=footnotesize]{subfig}

\usetikzlibrary{arrows.meta, calc}
\usepgfplotslibrary{colorbrewer}            

\pgfplotsset{
    compat=1.17,
    cycle list={{Paired-B},{Paired-A},{Paired-D}},
    every axis plot/.append style={
        thick
    }
}

\tikzstyle{agent}        = [draw, circle, fill=gray!30, inner sep=2pt]
\tikzstyle{hidden}       = [dashed, draw=black!80]
\tikzstyle{hidden agent} = [agent, hidden, fill=gray!15]

\newcommand{\bbma}{\begin{bmatrix}}
\newcommand{\ebma}{\end{bmatrix}}
\newcommand{\real}{\mathbb{R}}

\newtheorem{theorem}{Theorem}
\newtheorem{lemma}[theorem]{Lemma}

\theoremstyle{definition}
\newtheorem{assumption}{Assumption}
\newtheorem{definition}{Definition}

\theoremstyle{remark}
\newtheorem*{remark}{Remark}

\newcommand{\htwo}{{\ensuremath{\mathcal{H}_2}}}
\NewDocumentCommand{\name}{m}{\textsc{#1}}

\providecommand{\given}{}
\NewDocumentCommand{\renewgiven}{O{\delimsize}}{\renewcommand\given{\nonscript\:#1\vert\allowbreak\nonscript\:\mathopen{}}}

\DeclarePairedDelimiterX\set[1]\{\}{\renewgiven #1}
\DeclarePairedDelimiterXPP\prob[1]{\operatorname{\mathbb{P}}}(){}{\renewgiven #1}
\DeclarePairedDelimiterXPP\expect[1]{\operatorname{\mathbb{E}}}[]{}{\renewgiven #1}
\DeclarePairedDelimiterXPP\variance[1]{\operatorname{Var}}[]{}{\renewgiven #1}

\DeclarePairedDelimiterXPP\htwonorm[1]{}\|\|{_\htwo}{#1}
\DeclarePairedDelimiterXPP\diag[1]{\operatorname{diag}}(){}{#1}
\DeclarePairedDelimiterXPP\trace[1]{\operatorname{tr}}(){}{#1}
\DeclarePairedDelimiterXPP\indicator[2]{\operatorname{\mathbb{I}}_{#1}}(){}{#2}

\makeatletter
\renewcommand*\env@matrix[1][*\c@MaxMatrixCols c]{%
	\hskip -\arraycolsep
	\let\@ifnextchar\new@ifnextchar
	\array{#1}}
\newcommand*{\matline}[1]{\cmidrule(lr){1-#1}}
\makeatother

\newcommand{\arxivTF}[2]{#1}

\usepackage{fancyhdr}

\fancypagestyle{firstpage}{%
    \fancyfoot[C]{\small
        © 2023 IEEE.
        Personal use of this material is permitted.
        Permission from IEEE must be obtained for all other uses, in any current or future media, including reprinting/republishing this material for advertising or promotional purposes, creating new collective works, for resale or redistribution to servers or lists, or reuse of any copyrighted component of this work in other works.%
    }%
}

\title{\LARGE \bf
A Scalable Approach for Analysing Multi-Agent Systems with Heterogeneous Stochastic Packet Loss
}

\author{Christian Hespe and Herbert Werner%
\thanks{C. Hespe and H. Werner are with the Hamburg University of Technology, Institute of Control Systems, Eißendorfer Straße 40, 21073 Hamburg, Germany, {\tt\small \{christian.\allowbreak hespe, h.werner\}@tuhh.de}}%
}

\begin{document}

\maketitle
\thispagestyle{firstpage}

\begin{abstract}
    An important aspect in jointly analysing networked control systems and their communication is to model the networking in a sufficiently rich but at the same time mathematically tractable way.
As such, this paper improves on a recently proposed scalable approach for analysing multi-agent systems with stochastic packet loss by allowing for heterogeneous transmission probabilities and temporal correlation in the communication model.
The key idea is to consider the transmission probabilities as uncertain, which facilitates the use of tools from robust control.
Due to being formulated in terms of linear matrix inequalities that grow linearly with the number of agents, the result is applicable to very large multi-agent systems, which is demonstrated by numerical simulations with up to 10000 agents.

\end{abstract}

\pagestyle{empty}
\section{Introduction}\label{sec:intro}
The problem of controlling large-scale dynamical systems has attracted copious amounts of research activity in the past decades.
For almost all works on this subject, their scalability is of utmost importance and as such distributed and decentralized approaches offer distinctive advantages over centralized ones \cite{Massioni2009}.
In particular, this holds for multi-agent systems (MASs), in which small-scale dynamical systems -- called agents and typically only coupled through their mission -- collectively solve a control task, such as  distributed estimation or source seeking \cite{Mesbahi2010}.

In many MASs information is exchanged between agents using a communication network.
Such networks are, however, inherently unreliable and transmitted packets are \emph{not} guaranteed to be received.
As described in \cite{Schenato2007}, this \emph{random} loss of information can lead to a decrease in system performance and even loss of stability,
Nonetheless, communication effects are neglected in most works concerning networked MASs.

A suitable framework for modelling systems with random loss of information are Markov jump linear systems (MJLSs), which are switched linear systems whose switching is controlled by a Markov chain \cite{Costa2005}.
There is a rich literature on analysing MJLSs, both in terms of stability and system performance.
A direct application of the general analysis conditions does, however, lead to a combinatorial explosion for MASs with more than a few tens of agents and even applying simplifications as proposed in \cite{Wu2012} does not allow to consider truly large-scale MAS due to the remaining quadratic scaling of the computational complexity in terms of the number of agents.
For that reason, existing approaches on MASs with stochastic packet loss require additional assumptions on the packet loss model and are mostly concerned with stability analysis rather than system performance:
Many results rely on having \emph{identical} packet loss, that is, all links fail at the same time \cite{Ma2020}, which is a very restrictive assumption and violated in almost all real world scenarios.
Approaches that consider not identical \emph{loss} but homogeneous \emph{loss probability} are closer to reality, amongst which are \cite{Mesbahi2010, Patterson2010, Wu2012, Ghadami2012, Hespe2022}, all assuming Bernoulli distributed packet loss, i.e., no correlation of the packet loss across time steps.
\emph{Heterogeneous} loss probabilities are considered in only few previous works, e.g. \cite{Zhang2012} with tree graphs and uncertain probabilities and \cite{Ma2020} for arbitrary graphs and known probabilities.
Similarly, packet loss described by Markov chains, such that correlation in time can occur on individual links, is rarely studied, see, e.g., \cite{Xu2020} with identical packet loss on all links and \cite{Li2012} for a \emph{non-scalable} method with arbitrary and uncertain communication failures.

Only recently, \cite{Hespe2022} proposed to merge MJLSs with the decomposable systems framework \cite{Massioni2009}.
Under the assumption of Bernoulli distributed packet loss with homogeneous probability, this combination leads to \emph{linearly-scaling}, sufficient conditions for stability and performance that are applicable to MASs with thousands of agents and can be used for distributed controller synthesis \cite{Hespe2023}.
Building on these ideas, this paper extends the results to packet loss with heterogeneous transmission probabilities and limited correlation in time, while preserving the scalability of the decomposable systems approach.
This is achieved by considering the transmission probabilities as uncertain with \emph{uniform} bounds and involving tools from robust control.
Different from \cite{Zhang2012}, the proposed approach is not restricted to tree graphs and can in a straight-forward way be used for performance analysis.

After this introduction, the paper proceeds with detailing the agent and network model under consideration in Section~\ref{sec:mas}.
Section~\ref{sec:laplacians} introduces two supporting lemmata on calculating expected Laplacian matrices, which are used in Section~\ref{sec:robust} to derive the main robust analysis condition.
Finally, Section~\ref{sec:example} demonstrates the scalability of the results on a large consensus example before Section~\ref{sec:conclusions} concludes the paper.

\subsection{Contribution}\label{sec:intro_contribution}
The contribution of this paper is a linear matrix inequality (LMI) based condition to verify mean-square stability and bound the \htwo-norm from above.
In contrast to existing approaches, the condition is at the same time
\begin{enumerate}[i)]
    \item applicable to systems with heterogeneous packet loss probabilities,
    \item allows for correlation in time on individual links provided the packet loss is symmetric, and
    \item scales linearly with the number of agents.
\end{enumerate}
The result can be found in Theorem~\ref{thm:robust_h2}.

\subsection{Notation and Definitions}\label{sec:intro_notation}
In this paper, we denote the \(n \times n\) identity matrix as \(I_n\), dropping the index where it can be deduced from context.
The relation \(M \succ (\prec) \; 0\) means that \(M\) is symmetric positive (negative) definite and \(*\) indicates matrix blocks required for symmetry.
The Kronecker product is written as \(M_1 \otimes M_2\).
To denote the \emph{element wise} expectation and variance of the random matrix \(M\), we use \(\expect{M}\) and \(\variance{M}\), respectively.
Depending on \(z\), \(\|z\|\) stands for both the Euclidean vector norm and the 2-norm of stochastic signals defined by \(\|z\|^2 \coloneqq \sum_{k = 0}^\infty \expect{z^\top_k z_k}\).

A mathematical graph \(\mathcal{G} \coloneqq (\mathcal{V}, \mathcal{E})\) is composed of the vertex set~\(\mathcal{V} = \set{1, 2, \ldots, N}\) and the edge set~\(\mathcal{E} \subset \mathcal{V} \times \mathcal{V}\).
Defining \(e^{ij} \coloneqq (i, j)\), the edge \(e^{ij}\) is read as pointing from vertex~\(j\) to vertex~\(i\) and no self-loops exist, i.e., \(e^{ii} \notin \mathcal{E}\).
\(\mathcal{G}\) is called undirected if \(e^{ij} \in \mathcal{E} \Leftrightarrow e^{ji} \in \mathcal{E}\).
For each vertex \(i \in \mathcal{V}\), we define its in- and out-neighbourhood as \(\mathcal{N}_i^- \coloneqq \set{j \in \mathcal{V} \given e^{ij} \in \mathcal{E}}\) and \(\mathcal{N}_i^+ \coloneqq \set{j \in \mathcal{V} \given e^{ji} \in \mathcal{E}}\), respectively.
By inverting the direction of every edge, we obtain the transposed graph \(\mathcal{G}^\top = (\mathcal{V}, \mathcal{E}^\top)\), where \(\mathcal{E}^\top \coloneqq \set{e^{ij} \in \mathcal{V} \times \mathcal{V} \given e^{ji} \in \mathcal{E}}\).

\section{Multi-Agent Systems with Packet Loss}\label{sec:mas}
\subsection{Stochastic Jump Linear Systems}\label{sec:mas_mjls}
The concern of this paper are networked linear MASs which are subject to stochastic packet loss.
Due to the losses, their interconnection structure has to be considered switched rather than time-invariant, implying the same for the dynamics.
A fitting modelling framework to describe such systems are MJLSs.
For these switched linear systems, a Markov chain with state \(\sigma_k \in \mathcal{K} \coloneqq \set{1, 2, \ldots, m}\) determines which of the \(m\) modes is active at any given time~\(k\).
The dynamics of an MAS with \(N\) agents are then described by
\begin{equation}\label{eq:mjls}
    G: \,\left\{\enspace\begin{aligned}
        x_{k+1} &= A_{\sigma_k} x_k + B_{\sigma_k} w_k, \\
        z_k     &= C_{\sigma_k} x_k + D_{\sigma_k} w_k,
    \end{aligned}\right.
\end{equation}
where \(x_k \in \real^{N n_x}\) is the dynamic state of the system and \(w_k \in \real^{N n_w}\) and \(z_k \in \real^{N n_z}\) are the performance input and output, respectively.
On the other hand, the Markov chain evolves in time according to the transition probabilities
\begin{equation}\label{eq:mjls_probability}
    \prob[\big]{\sigma_{k+1} = j \given \sigma_k = i} = t_{ij}
\end{equation}
and the initial distribution is \(\prob{\sigma_0 = i} = \mu_i\).

Because of the stochastic switching, considering asymptotic stability of the MJLS~\(G\) in the usual deterministic sense is often more restrictive than desired or achievable.
Instead, this paper will focus on asymptotic mean-square stability.
\begin{definition}[Mean-Square Stability \cite{Costa2005}]\label{def:stability}
    The MJLS~\eqref{eq:mjls} is mean-square stable if
    \begin{align*}
        \lim\limits_{k \to \infty} \expect[\big]{\|x(k)\|} = 0 & &
        \text{and} & &
        \lim\limits_{k \to \infty} \expect[\big]{\|x(k) x^\top(k)\|} = 0
    \end{align*}
    for all initial conditions \(x_0\) and \(\sigma_0\).
\end{definition}

Apart from mean-square stability, there exists a variety of other stochastic stability definitions, e.g. stability in probability and almost sure stability.
A particularly attractive property of mean-square stability is however the existence of stability tests in the form of LMIs.
Furthermore, mean-square stability implies stability as in the other two definitions \cite{Costa2005}.

\begin{theorem}[Stability Test \cite{Costa1993}]\label{thm:mjls_stability}
    The MJLS~\eqref{eq:mjls} is mean-square stable if and only if there exist \(X_i \succ 0\) such that
    \begin{equation}\label{eq:ms_stability_lmi}
        \sum_{j \in \mathcal{K}} t_{ij} A_j^\top X_j A_j - X_i \prec 0 \qquad \forall\, i \in \mathcal{K}.
    \end{equation}
\end{theorem}

The stability test in Theorem~\ref{thm:mjls_stability} can be evaluated efficiently as a semidefinite program as long as the number of modes \(m\) is sufficiently small.
This will however \emph{not} be the case for the MASs considered in the following sections.

In addition to stability, we consider system performance in terms of the MJLS \htwo-norm:

\begin{definition}[MJLS \htwo-Norm \cite{Costa2005}]\label{def:h2}
    The \htwo-norm of the MJLS~\eqref{eq:mjls} is defined as
    \begin{equation*}
        \htwonorm{G}^2 \coloneqq \sum_{i \in \mathcal{K}} \sum_{s = 1}^{N n_w} \mu_i \big\|z^{s,i}\big\|^2,
    \end{equation*}
    where \(z^{s,i}\) is the response of \(G\) to a discrete impulse applied at the \(s\)th input with \(\sigma_0 = i\) and \(x_0 = 0\).
\end{definition}

Like for mean-square stability, there exist efficient means to calculate the \htwo-norm in terms of LMIs for small $m$.
The following theorem is an equivalent reformulation of the conditions that originally appeared in \cite{Costa1997}, which is more suitable for the analysis in the remainder of this paper.
\begin{theorem}[\htwo-Norm Calculation \cite{Fioravanti2012}]\label{thm:mjls_h2}
    Given the stable MJLS~\eqref{eq:mjls}, \(\htwonorm{G} < \gamma\) if and only if there exist \(X_i \succ 0\) and a symmetric \(Z\) with \(\trace{Z} < \gamma^2\) such that%
    \begin{subequations}\label{eq:mjls_h2_lmi}\begin{align}%
        \sum_{j \in \mathcal{K}} t_{ij} \left(A_j^\top X_j A_j + C_j^\top C_j\right) - X_i &\prec 0, \label{eq:mjls_h2_gramian_lmi} \\
        \sum_{j \in \mathcal{K}} \mu_j  \left(B_j^\top X_j B_j + D_j^\top D_j\right) - Z   &\prec 0, \label{eq:mjls_h2_trace_lmi}
    \end{align}\end{subequations}
    for all \(i \in \mathcal{K}\).
\end{theorem}

\subsection{Modelling Packet Loss using Stochastic Laplacians}\label{sec:mas_loss}
In order to model the exchange of information in the networked MAS we are using tools from graph theory.
The \emph{nominal} interconnection structure, i.e., if no packet loss occurs, is described by the graph \(\mathcal{G}^0 = (\mathcal{V}, \mathcal{E}^0)\), where there is a one-to-one correspondence between agents and vertices.
In the dynamic equations of the MAS, the graph will appear through its Laplacian \(L(\mathcal{G}) \in \mathbb{Z}^{N \times N}\), which is defined element wise as \(L(\mathcal{G}) \coloneqq [l_{ij}(\mathcal{G})]\) with
\begin{equation}\label{eq:laplacian_nominal}
    l_{ij}(\mathcal{G}) \coloneqq \begin{cases}
        -1   & if \(i \neq j\) and \(e^{ij} \in \mathcal{E}\), \\
        0   & if \(i \neq j\) and \(e^{ij} \notin \mathcal{E}\), \\
        -\textstyle{\sum_{l \neq i}} l_{il}(\mathcal{G}) & if \(i = j\).
    \end{cases}
\end{equation}

To capture the loss of packets, introduce stochastic processes \(\{\theta_k^{ij}\}\) for all \(e^{ij} \in \mathcal{E}^0\).
The processes take values in \(\set{0,1}\), where \(\theta_k^{ij} = 1\) means that information is successfully transmitted on \(e^{ij}\) at step~\(k\) and \(\theta_k^{ij} = 0\) that information is lost.
In this paper, we assume all \(\{\theta_k^{ij}\}\) are described by Markov chains, which are independent if the processes belong to distinct pairs of agents.
\begin{assumption}\label{ass:markov}
    The processes \(\{\theta_k^{ij}\}\) are described by Markov chains such that, for all \(k \geq 0\) and \(e^{ij} \in \mathcal{E}^0\), we have
    \begin{subequations}\label{eq:markov_probability}\begin{align}
        \prob[\big]{\theta_{k+1}^{ij} = 1 \given \theta_k^{ij} = 1} &= p^{ij}, \label{eq:markov_probabilits_success}\\
        \prob[\big]{\theta_{k+1}^{ij} = 1 \given \theta_k^{ij} = 0} &= q^{ij}, \label{eq:markov_probabilits_failure}\\
        \prob[\big]{\theta_0^{ij} = 1} &= \eta^{ij}. \label{eq:markov_probabilities_initial}
    \end{align}\end{subequations}
    Furthermore, \(\theta_k^{ij}\) and \(\theta_{k'}^{rs}\) are independent random variables for all \(k, k' \geq 0\) whenever \(e^{ij} \neq e^{rs} \neq e^{ji}\).
\end{assumption}

Mapping between \(\{\theta_k^{ij}\}\) and \(\{\sigma_k\}\) can be accomplished by treating \(\theta_k^{ij}\) as the digits of a binary representation of \(\sigma_k\).
This shows that the number of modes is \(m = 2^{\lvert\mathcal{E}^0\rvert}\).
Additionally, introduce \(\mathcal{G}_i = (\mathcal{V}, \mathcal{E}_i)\), where \(\mathcal{E}_i \subseteq \mathcal{E}^0\) is the set of edges that successfully transmit information with \(\sigma_k = i\).
The Laplacian \(\{L(\mathcal{G}_{\sigma_k})\}\) is then a stochastic process, describing the exchange of information subject to packet loss.
In the following, we use the shorthand notations \(L^0 \coloneqq L(\mathcal{G}^0)\) and \(L_i \coloneqq L(\mathcal{G}_i)\) if the graph can be deduced from context.

\subsection{Decomposable Markov Jump Linear Systems}\label{sec:mas_decomposable}
Due to the growth in computational complexity for large \(N\) in Theorems~\ref{thm:mjls_stability} and \ref{thm:mjls_h2}, analysing the MJLS~\eqref{eq:mjls} in its full generality is only tractable for MASs with small to moderate size.
For MASs with homogeneous linear time-invariant agents and ideal communication, the decomposable systems framework proposed in \cite{Massioni2009} offers a flexible approach to alleviate this issue.
The approach relies on system matrices that are structured as \(A = I_N \otimes A^d + S \otimes A^c\), where \(A^d\) and \(A^c\) are the decoupled and coupled components, respectively, and \(S\) is called the pattern matrix.
It is then possible to selectively apply a similarity transformation to the pattern matrix and, for diagonalizable~\(S\), decouple the MAS into \(N\) subsystems with the size of a single agent.

Based on this idea, \cite{Hespe2022, Hespe2023} proposed to consider \emph{decomposable} MJLSs with system matrix
\begin{equation}\label{eq:matrix_structure}
    A_i = I_N \otimes A^d + L_i \otimes A^c + L^0 \otimes A^p,
\end{equation}
and analogously for \(B_i\), \(C_i\) and \(D_i\).
In these MJLS, a third component intended for deterministic performance channels is introduced in addition to the coupled and decoupled parts.
Decoupling the MAS directly as outlined above is impossible because of the time-varying pattern matrix \(L_{\sigma_k}\).
Nonetheless, it was shown in \cite{Hespe2022} that -- under certain assumptions on the packet loss -- the same transformation can be applied to decouple LMI-based stability and performance tests.
Utilizing that approach with the packet loss model described in Assumption~\ref{ass:markov} is, however, posed to fail.
This paper will thus develop an alternative strategy for decomposition based on concepts from robust control.

\section{Expected Laplacians with Heterogeneous Probabilities}\label{sec:laplacians}
A key ingredient to the approach proposed in \cite{Hespe2022} is to analytically calculate the expectation of the Laplacian \(L_{\sigma_k}\) and the product \(L_{\sigma_k}^\top L_{\sigma_k}\).
Similar calculations were performed before in \cite{Wu2012} without exposing the inherent structure of the result, which is required for decoupling.
However, neither result is applicable to stochastic Laplacians as resulting from Assumption~\ref{ass:markov} because both require homogeneous loss probabilities and independence in time, i.e., \(p^{ij} = q^{ij} = p\) for all \(e^{ij} \in \mathcal{E}^0\).
We thus generalize the existing results to heterogeneous probabilities in the following lemma:
\begin{lemma}\label{lem:heterogeneous_expectation}
    Given the graph~\(\mathcal{G}^0\) and packet loss satisfying Assumption~\ref{ass:markov}, the conditional expectation of the Laplacian product is
    \begin{multline}\label{eq:heterogeneous_expectation}
        \expect[\big]{L_{\sigma_{k+1}}^\top L_{\sigma_{k+1}} \given \sigma_k} =
        \expect[\big]{L_{\sigma_{k+1}} \given \sigma_k}^\top \expect[\big]{L_{\sigma_{k+1}} \given \sigma_k} \\
        + \variance[\big]{L(\mathcal{G}_{\sigma_{k+1}}) \given \sigma_k} + \variance[\big]{L(\mathcal{G}_{\sigma_{k+1}}^\top) \given \sigma_k}.
    \end{multline}
\end{lemma}
\begin{proof}
    \arxivTF{The \((i,j)\)-entry of \(\expect[\big]{L_{\sigma_{k+1}}^\top L_{\sigma_{k+1}} \given \sigma_k}\) is
\begin{equation*}
    \sum_{s=1}^N \expect[\big]{l_{si}(\mathcal{G}_{\sigma_{k+1}}) l_{sj}(\mathcal{G}_{\sigma_{k+1}}) \given \sigma_k}
    \eqqcolon \sum_{s=1}^N \beta_{ij}^s.
\end{equation*}
By doing a five-fold case distinction in \(i\), \(j\) and \(s\), \(\beta_{ij}^s\) can be calculated as
\begin{align*}
    \beta_{ii}^i &= \textstyle \sum_{t \in \mathcal{N}_i^-} \sum_{r \in \mathcal{N}_i^-} \alpha_k^{it} \alpha_k^{ir} + \sum_{r \in \mathcal{N}_i^-} \alpha_k^{ir}\big(1- \alpha_k^{ir}\big), \\
    \beta_{ii}^s &= \alpha_k^{si} \indicator{\mathcal{N}_i^+}{s}, \\
    \beta_{ij}^i &= \textstyle -\left[\sum_{r \in \mathcal{N}_i^-} \alpha_k^{ir} \alpha_k^{ij} + \alpha_k^{ij}\big(1-\alpha_k^{ij}\big)\right] \indicator{\mathcal{N}_i^-}{j}, \\
    \beta_{ij}^j &= \textstyle -\left[\sum_{r \in \mathcal{N}_j^-} \alpha_k^{ji} \alpha_k^{jr} + \alpha_k^{ji}\big(1-\alpha_k^{ji}\big)\right] \indicator{\mathcal{N}_j^-}{i}, \\
    \beta_{ij}^s &= \alpha_k^{si} \alpha_k^{sj} \indicator{\mathcal{N}_i^+ \cap \mathcal{N}_j^+}{s},
\end{align*}
where \(\indicator{S}{x}\) is the set-membership indicator function
\begin{equation*}
    \indicator{S}{x} \coloneqq \begin{cases}
        1 & if \(x \in S\), \\
        0 & else
    \end{cases}
\end{equation*}
and \(\alpha_k^{ij} \coloneqq \expect[\big]{\theta_{k+1}^{ij} \given \theta_k^{ij}}\).
Summing over \(s\), we obtain
\begin{multline*}
    \sum_{s = 1}^N \beta_{ii}^s - \expect[\big]{l_i(\mathcal{G}_{\sigma_{k+1}}) \given \sigma_k}^\top \expect[\big]{l_i(\mathcal{G}_{\sigma_{k+1}}) \given \sigma_k} =\\
    \textstyle \sum_{r \in \mathcal{N}_i^-} \alpha_k^{ir}\big(1- \alpha_k^{ir}\big) + \sum_{s \in \mathcal{N}_i^+} \alpha_k^{si}\big(1- \alpha_k^{si}\big)
\end{multline*}
for the diagonal entries and
\begin{multline*}
    \sum_{s = 1}^N \beta_{ij}^s - \expect[\big]{l_i(\mathcal{G}_{\sigma_{k+1}}) \given \sigma_k}^\top \expect[\big]{l_j(\mathcal{G}_{\sigma_{k+1}}) \given \sigma_k} =\\
    -\alpha_k^{ij}\big(1-\alpha_k^{ij}\big) \indicator{\mathcal{N}_i^-}{j} - \alpha_k^{ji}\big(1-\alpha_k^{ji}\big) \indicator{\mathcal{N}_j^-}{i}
\end{multline*}
for the off-diagonal ones.
Noting that \(\variance[\big]{\theta_{k+1}^{ij} \given \theta_k^{ij}} = \alpha_k^{ij} \big(1-\alpha_k^{ij}\big)\), these correspond to the element wise variance of \(L\big(\mathcal{G}_{\sigma_k}\big)\) and \(L\big(\mathcal{G}_{\sigma_k}^\top\big)\), respectively.
}{Omitted for brevity, see \cite{ExtendedVersion}.}
\end{proof}

Lemma~\ref{lem:heterogeneous_expectation} shows that the expectation and variance of the Laplacian can be evaluated element wise before calculating the matrix product.
However, the eigenvectors of \(\expect{L_{\sigma_{k+1}}^\top L_{\sigma_{k+1}} \given \sigma_k}\) will generally not be the same as those of \(\expect{L_{\sigma_{k+1}} \given \sigma_k}\), thus they cannot be used for decomposing the LMIs.
For that reason, we introduce additional assumptions that make an alternative factorization viable:
\begin{assumption}\label{ass:symmetric}
    The communication graph \(\mathcal{G}^0\) is undirected and the transition probabilities are symmetric, i.e. \(p^{ij} = p^{ji}\), \(q^{ij} = q^{ji}\), and \(\eta^{ij} = \eta^{ji}\) for all \(e^{ij} \in \mathcal{E}^0\).
    Furthermore, the packet loss is either
    \begin{enumerate}[i)]
        \item symmetric, i.e. \(\theta_k^{ij} = \theta_k^{ji}\) for all \(k \geq 0\), \(e^{ij} \in \mathcal{E}^0\), or \label{ass:symmetric_markov}
        \item independent in time, i.e. \(p^{ij} = q^{ij}\) for all \(e^{ij} \in \mathcal{E}^0\). \label{itm:ass_symmetric_bernoulli}
    \end{enumerate}
\end{assumption}

In the following, let \(\Phi \in \set{{-1}, 0, 1}^{N \times \lvert\mathcal{E}^0\rvert}\) denote the incidence matrix \cite{Mesbahi2010} of the nominal graph, such that \(L^0 = \Phi \Phi^\top\).
The incidence matrix has one column for each \(e^{ij} \in \mathcal{E}^0\), where the \(i\)th entry of that column is 1, the \(j\)th is -1, and all others are 0.
Furthermore, we use \(\Theta_k\) for the matrix obtained by diagonal concatenation of \(\theta_k^{ij}\) in matching order to the columns of \(\Phi\).
We obtain the following factorizations:

\begin{lemma}\label{lem:symmetric_expectation}
    Given the graph~\(\mathcal{G}^0\) and packet loss satisfying Assumptions~\ref{ass:markov} and \ref{ass:symmetric}, the conditional expectation and variance of the Laplacian are given by
    \begin{align}
        \expect[\big]{L_{\sigma_{k+1}} \given \sigma_k}   &= \Phi \expect[\big]{\Theta_{k+1} \given \sigma_k} \Phi^\top, \label{eq:symmetric_expectation}\\
        \variance[\big]{L_{\sigma_{k+1}} \given \sigma_k} &= \Phi \variance[\big]{\Theta_{k+1} \given \sigma_k} \Phi^\top. \label{eq:symmetric_variance}
    \end{align}
\end{lemma}
\begin{proof}
    The independence assumption of \(\theta_k^{ij}\) in Assumption~\ref{ass:markov} ensures that the expectation and variance can be applied to each \(\theta_k^{ij}\) individually, hence both are weighted variants of \(L^0\), where the weights are the respective stochastic moments of the edges.
    Moreover, Assumption~\ref{ass:symmetric} guarantees that \(\expect{L_{\sigma_{k+1}} \given \sigma_k}\) and \(\variance{L_{\sigma_{k+1}} \given \sigma_k}\) are symmetric for all \(\sigma_k\).
    Therefore, we obtain \eqref{eq:symmetric_expectation} and \eqref{eq:symmetric_variance} as factorizations of symmetric weighted Laplacians \cite{Mesbahi2010}.
\end{proof}

The core feature of the factorizations in \eqref{eq:symmetric_expectation} and \eqref{eq:symmetric_variance} is that they separate the nominal communication structure of the MAS -- which is described by \(\Phi\) -- from the effects induced by packet loss in \(\Theta_k\).

\section{Analysis with Uncertain Probabilities}\label{sec:robust}
\subsection{Robust Stability}\label{sec:robust_stability}
The analytic calculation of the expected Laplacians above can now be used to obtain scalable sufficient conditions for stability and \htwo-performance.
As a first step, notice that \eqref{eq:ms_stability_lmi} is equivalent to
\begin{equation}\label{eq:stability_expectation}
    \expect[\big]{A_{\sigma_{k+1}}^\top X_{\sigma_{k+1}} A_{\sigma_{k+1}} \given \sigma_k} - X_{\sigma_k} \prec 0
\end{equation}
for all \(k \geq 0\).
Inserting \eqref{eq:matrix_structure} and expanding the product, only \(\expect{X_{\sigma_{k+1}} \given \sigma_k}\), \(\expect{X_{\sigma_{k+1}} (L_{\sigma_{k+1}} \otimes I_{n_x}) \given \sigma_k}\), and \[\expect[\big]{(L_{\sigma_{k+1}} \otimes I_{n_x})^\top X_{\sigma_{k+1}} (L_{\sigma_{k+1}} \otimes I_{n_x}) \given \sigma_k}\] have to be evaluated, because all other terms are constant and can be pulled out of the expectation.
Moreover, by enforcing that all \(X_i\) are identical and block-diagonal with repeated blocks, i.e., \(X_i = X_j = I_N \otimes Y\) for all \(i,j \in \mathcal{K}\), and utilizing the commutation property \((M_1 \otimes I)(I \otimes M_2) = (I \otimes M_2)(M_1 \otimes I)\), we can pull out \(I_N \otimes Y\) from the expectations, such that Lemma~\ref{lem:heterogeneous_expectation} is applicable.
Thus, the existence of \(Y \succ 0\) such that
\begin{multline}\label{eq:stability_sufficient}
    \expect[\big]{A_{\sigma_{k+1}} \given \sigma_k}^\top (I_N \otimes Y) \expect[\big]{A_{\sigma_{k+1}} \given \sigma_k} - I_N \otimes Y \\
    + 2\variance[\big]{L_{\sigma_{k+1}} \given \sigma_k} \otimes \big(A^{c\top} Y A^c\big) \prec 0
\end{multline}
holds, is a sufficient condition for \eqref{eq:ms_stability_lmi}.
Finally, \eqref{eq:stability_sufficient} can be factorized as
\begin{multline}\label{eq:stability_factorized}
    \bbma * \\ * \\ * \ebma^\top
    \bbma
        -I_N \otimes Y & & \\
         & I_N \otimes Y & \\
         & & 2 I_m \otimes Y
    \ebma \\
    \cdot \bbma
        I_N \otimes I_{n_x} \\
        \expect[\big]{A_{\sigma_{k+1}} \given \sigma_k} \\
        \big(\variance[\big]{\Theta_{k+1} \given \sigma_k}^{\frac 12} \Phi^\top\big) \otimes A^c
    \ebma \prec 0
\end{multline}
by applying Lemma~\ref{lem:symmetric_expectation}.
Note that while the blocks of the centre matrix are decoupled, the agents are coupled through the Laplacian that appears in the outer factors.

\begin{remark}
    Imposing that \(X_i = X_j\) for all \(i,j \in \mathcal{K}\) is in general a conservative choice required for applying Lemma~\ref{lem:heterogeneous_expectation}.
    However, in case Assumption~\ref{ass:symmetric} case~\ref{itm:ass_symmetric_bernoulli} holds, this restriction can be imposed without loosing necessity, cf. \cite{Fioravanti2012}.
    In either case, further conservatism is incurred from imposing that \(X\) has block-repeated structure, i.e., \(X = I_N \otimes Y\).
\end{remark}

In \cite{Hespe2022}, it is proposed to apply a diagonalizing transformation to an equivalent of \eqref{eq:stability_factorized} to obtain a decoupled set of LMIs.
However, the transformation relies on the assumption of packet loss that is independent in time with \emph{homogeneous} transmission probabilities and is therefore not applicable.
Instead, we propose to consider the transition probabilities of each edge as uncertain with \emph{uniform} upper and lower bounds:
\begin{assumption}\label{ass:bounded_loss}
    There exist constants \(0 \leq \rho_l \leq \rho_u \leq 1\) such that \(p^{ij}, q^{ij}, \eta^{ij} \in [\rho_l, \rho_u]\) for all \(e^{ij} \in \mathcal{E}^0\).
\end{assumption}

A powerful and flexible tool from robust control to handle parametric uncertainties are linear fractional transformations (LFTs) in combination with the full block S-procedure (FBSP) \cite{Scherer2000}.
In spite of a non-rational dependence on the uncertainty in \eqref{eq:stability_factorized}, it can be represented in LFT form by embedding the uncertainty in a higher dimensional space.
To this end, introduce new variables \(a_{ij}, b_{ij} \geq 0\) with
\begin{equation*}
    a_{ij}^2 \coloneqq \expect[\big]{\theta_{k+1}^{ij} \given \theta_k^{ij}}
    \qquad\text{and}\qquad
    b^2 \coloneqq 1 - a^2.
\end{equation*}
Together, we obtain
\begin{equation*}
    \sqrt{\variance[\big]{\theta_{k+1}^{ij} \given \theta_k^{ij}}} = a_{ij} b_{ij},
\end{equation*}
such that \eqref{eq:stability_factorized} depends polynomially on the uncertainty in terms of \(a_{ij}\) and \(b_{ij}\).
The uncertainty set of the LFT for a single link can then be described by the \emph{non-convex} set
\begin{equation}\label{eq:uncertainty_set}
    \bm{\Delta} \coloneqq \set*{
        \bbma
            a & 0 \\
            b & 0 \\
            0 & a
        \ebma
        \given
        \begin{aligned}
            a, b &\geq 0               \\[-0.5ex]
            a^2  &\in [\rho_l, \rho_u] \\[-0.5ex]
            b^2  &= 1 - a^2
        \end{aligned}
    }.
\end{equation}
Additionally, the FBSP requires the so-called multiplier set
\begin{equation}\label{eq:multiplier_set}
    \mathcal{P}_\alpha \coloneqq \set*{P = P^\top \given \bbma * \\ * \ebma^\top \!\! P \bbma \Delta \otimes I_\alpha \\ I_{2\alpha} \ebma \succ 0 \enspace \forall \Delta \in \bm{\Delta}},
\end{equation}
where \(\alpha\) denotes the dimension of the information exchanged amongst agents.
At last, applying the FBSP to \eqref{eq:stability_factorized} leads to the following robust stability test:

\begin{figure*}[!t]
    \normalsize

    \newcounter{storeeqnno}
    \setcounter{storeeqnno}{\value{equation}}
    \setcounter{equation}{16}

    \setlength\arraycolsep{3pt}
    \begin{subequations}\begin{gather}
        \addtocounter{equation}{2}
        \bbma
            I & 0 \\
            A^d + \lambda_i A^p & \multirow{2}{*}{$\sqrt{\lambda_i}\mathcal{B}_1$} \\
            C^d + \lambda_i C^p & \\
            0 & \multirow{2}{*}{$\mathcal{B}_2$} \\
            0 &
        \ebma^\top \!\! \bbma
            -Y & & & & \\
            & Y & & & \\
            & & I & & \\
            & & & 2Y & \\
            & & & & 2I
        \ebma
        \bbma
            I & 0 \\
            A^d + \lambda_i A^p & \multirow{2}{*}{$\sqrt{\lambda_i}\mathcal{B}_1$} \\
            C^d + \lambda_i C^p & \\
            0 & \multirow{2}{*}{$\mathcal{B}_2$} \\
            0 &
        \ebma + \bbma
            0 & I \\
            0 & \multirow{2}{*}{$\mathcal{B}_3$} \\
            0 & \\
            \sqrt{\lambda_i} A^c & 0 \\
            \sqrt{\lambda_i} C^c & 0
        \ebma^\top \!\! P_1 \bbma
            0 & I \\
            0 & \multirow{2}{*}{$\mathcal{B}_3$} \\
            0 & \\
            \sqrt{\lambda_i} A^c & 0 \\
            \sqrt{\lambda_i} C^c & 0
        \ebma \prec 0 \label{eq:robust_h2_lmi_gramian_modal} \\
        \bbma
            I & 0 \\
            B^d + \lambda_i B^p & \multirow{2}{*}{$\sqrt{\lambda_i}\mathcal{B}_1$} \\
            D^d + \lambda_i D^p & \\
            0 & \multirow{2}{*}{$\mathcal{B}_2$} \\
            0 &
        \ebma^\top \!\! \bbma
            -Z_i & & & & \\
            & Y & & & \\
            & & I & & \\
            & & & 2Y & \\
            & & & & 2I
        \ebma
        \bbma
            I & 0 \\
            B^d + \lambda_i B^p & \multirow{2}{*}{$\sqrt{\lambda_i}\mathcal{B}_1$} \\
            D^d + \lambda_i D^p & \\
            0 & \multirow{2}{*}{$\mathcal{B}_2$} \\
            0 &
        \ebma + \bbma
            0 & I \\
            0 & \multirow{2}{*}{$\mathcal{B}_3$} \\
            0 & \\
            \sqrt{\lambda_i} B^c & 0 \\
            \sqrt{\lambda_i} D^c & 0
        \ebma^\top \!\! P_2 \bbma
            0 & I \\
            0 & \multirow{2}{*}{$\mathcal{B}_3$} \\
            0 & \\
            \sqrt{\lambda_i} B^c & 0 \\
            \sqrt{\lambda_i} D^c & 0
        \ebma \prec 0 \label{eq:robust_h2_lmi_trace_modal}
    \end{gather}\end{subequations}

    \hrulefill

    \setcounter{equation}{\value{storeeqnno}}
    \def\eqno{\relax}
\end{figure*}

\begin{theorem}[Robust Stability Test]\label{thm:robust_stability}
    Given a decomposable MJLS, a communication graph~\(\mathcal{G}^0\), and packet loss satisfying Assumptions~\ref{ass:markov} to \ref{ass:bounded_loss}, the MJLS is mean-square stable if there exist \(Y \succ 0\) and \(P \in \mathcal{P}_{n_x}\) such that
    \begin{subequations}\label{eq:robust_stability_lmi}\begin{gather}
        \allowdisplaybreaks
        A^{d\top} Y A^d - Y \prec 0 \label{eq:robust_stability_lmi_nominal}\\
        \bbma * \\ * \\ * \\ \matline{1} * \\ * \\ * \ebma^\top \!\!
        \bbma[ccc|c]
            -Y & & & \\
            & Y & & \\
            & & 2Y & \\ \matline{4}
            & & & P
        \ebma
        \bbma
            I & 0 \\
            A^d + \lambda_i A^p & \sqrt{\lambda_i} \mathcal{B}_1 \\
            0 & \mathcal{B}_2 \\ \matline{2}
            0 & I \\
            0 & \mathcal{B}_3 \\
            \sqrt{\lambda_i} A^c  & 0
        \ebma \prec 0 \label{eq:robust_stability_lmi_modal}
    \end{gather}\end{subequations}
    hold for all non-zero \(\lambda_i\), where \(\lambda_i\) are the eigenvalues of \(L^0\) and \(\mathcal{B}_j\) is the \(j\)th block-row of \(I_3 \otimes I_{n_x}\)
\end{theorem}
\begin{proof}
    \arxivTF{The idea behind Theorem~\ref{thm:robust_stability} is to use the FBSP to find a \(Y \succ 0\) that -- independent of \(\sigma_k\) -- satisfies \eqref{eq:stability_factorized} for all \(p^{ij}\), \(q^{ij}\) adhering to Assumptions~\ref{ass:markov} to \ref{ass:bounded_loss}.
To describe the uncertainty in the transition probabilities, we thus reformulate the outer factors of \eqref{eq:stability_factorized} in terms of the LFT
\begin{equation*}\begin{split}
    &\bbma
        \expect[\big]{A_{\sigma_{k+1}} \given \sigma_k} \\
        \big(\variance[\big]{\Theta_{k+1} \given \sigma_k}^{\frac 12} \Phi^\top\big) \otimes A^c
    \ebma = \\
    &\qquad\bbma
        I_N \otimes A^d + L^0 \otimes A^p \\
        0
    \ebma +
    \bbma
        \Phi \otimes \bbma I_{n_x} & 0 & 0 \ebma \\
        I_m \otimes \bbma 0 & I_{n_x} & 0 \ebma
    \ebma \\
    &\quad\cdot \left(I - \bar{\Delta} \left(I_m \otimes \bbma
        0 & 0 & I_{n_x} \\
        0 & 0 & 0
    \ebma \right)\right)^{-1}
    \bar{\Delta} \left(\Phi^\top \otimes \bbma
        0 \\ A^c
    \ebma\right),
\end{split}\end{equation*}
where \(\bar{\Delta}\) is given by a diagonal concatenation of \(\Delta^{ij}\) for all \(e^{ij} \in \mathcal{E}^0\) and \(\Delta^{ij}\) are
\begingroup
    \setlength\arraycolsep{2pt}
    \begin{align*}
        \Delta^{ij} = \bbma
            \sqrt{p^{ij}}   & 0 \\
            \sqrt{1-p^{ij}} & 0 \\
            0 & \sqrt{p^{ij}}
        \ebma
        & & \text{or} & &
        \Delta^{ij} = \bbma
            \sqrt{q^{ij}}   & 0 \\
            \sqrt{1-q^{ij}} & 0 \\
            0 & \sqrt{q^{ij}}
        \ebma
    \end{align*}
\endgroup
if \(\theta_k^{ij} = 1\) or \(\theta_k^{ij} = 0\), respectively.
The uncertainty is therefore captured by allowing \(\Delta^{ij}\) to be arbitrary in \(\bm{\Delta}\).
In the following, we use the notation \(\hat{M}\) for some matrix \(M\) to denote \(I_N \otimes M\).
We obtain
\begin{equation*}\begin{split}
    \bbma * \\ * \\ * \ebma^\top
    \bbma
        -\hat{Y} & & \\
        & \hat{Y} & \\
        & & I_m \otimes 2Y
    \ebma
    \bbma
        I & 0 \\
        \hat{A}^d + L^0 \otimes A^p & \Phi \otimes \mathcal{B}_1 \\
        0 & I_m \otimes \mathcal{B}_2
    \ebma& \\
    + \bbma * \\ * \ebma^\top \!\!
    \bar{P}
    \bbma
        0 & I \\
        \Phi^\top \otimes \bbma 0 \\ A^c \ebma & I_m \otimes \bbma \mathcal{B}_3 \\ 0 \ebma
    \ebma \prec 0&
\end{split}\end{equation*}
by applying the FBSP, where \(\bar{P}\) is the multiplier capturing the network-wide uncertainty \(\bar{\Delta}\).
In light of the goal to decompose the analysis conditions, we impose
\begin{align*}
    \bar{P} = \bbma
        I_m \otimes Q      & I_m \otimes S \\
        I_m \otimes S^\top & I_m \otimes R
    \ebma
    & & \text{and define} & &
    P \coloneqq \bbma
        Q      & S \\
        S^\top & R
    \ebma,
\end{align*}
where \(Q = Q^\top \in \real^{3n_x \times 3n_x}\) and \(R = R^\top \in \real^{2n_x \times 3n_x}\).

Take the singular value decomposition \(\Phi = U \Sigma V^\top\) and note that \(L^0 = U \Lambda U^\top\).
Then, apply a congruence transformation by multiplying with \(\diag{U \otimes I_{n_x}, V \otimes I_{3n_x}}\) and its transpose from right and left, respectively, and obtain
\begin{equation*}\begin{split}
    \bbma * \\ * \\ * \ebma^\top
    \bbma
        -\hat{Y} & & \\
        & \hat{Y} & \\
        & & I_m \otimes 2Y
    \ebma
    \bbma
        I & 0 \\
        \hat{A}^d + \Lambda \otimes A^p & \Sigma \otimes \mathcal{B}_1 \\
        0 & I_m \otimes \mathcal{B}_2
    \ebma& \\
    + \bbma * \\ * \ebma^\top \!\!
    \bar{P}
    \bbma
        0 & I \\
        \Sigma^\top \otimes \bbma 0 \\ A^c \ebma & I_m \otimes \bbma \mathcal{B}_3 \\ 0 \ebma
    \ebma \prec 0&.
\end{split}\end{equation*}
Note that each block is in diagonal form such that the inequality can be permuted into smaller decoupled LMIs.
The entries corresponding to the non-zero eigenvalues of \(L^0\) -- in both \(\Lambda\) and \(\Sigma\) --  lead to \eqref{eq:robust_stability_lmi_modal} and for \(\lambda_i = 0\) we obtain \eqref{eq:robust_stability_lmi_nominal}.
In addition, if \(m \geq N\), the remaining entries from the second block-column in the previous LMI result in repetitions of
\begin{equation*}
    2 \mathcal{B}_2^\top P \mathcal{B}_2 + \bbma
        I \\ \mathcal{B}_3 \\ 0
    \ebma^\top \!\! P \bbma
        I \\ \mathcal{B}_3 \\ 0
    \ebma \prec 0,
\end{equation*}
which is implied by \eqref{eq:robust_stability_lmi_modal} and thus redundant.
Finally, the multiplier condition on \(\bar{P}\) can be permuted into \(m\) identical copies in the form of \eqref{eq:multiplier_set}.
}{Omitted for brevity, see \cite{ExtendedVersion}.}
\end{proof}
\begin{remark}
    The inclusion \(P \in \mathcal{P}_{n_x}\) is described by an infinite set of LMIs and as such it is non-trivial to obtain a computationally tractable formulation.
    It is therefore customary to rely on approximations of \(\mathcal{P}_{n_x}\) for numerical evaluation.
    Amongst the possible techniques are gridding-based methods, D- and D/G-scalings, or sum-of-squares relaxations \cite{Hoffmann2015, Veenman2016}.
\end{remark}

Theorem~\ref{thm:robust_stability} paves the way for testing mean-square stability of large MAS with heterogeneous transmission probabilities in a scalable way.
The size of \(Y\) and \(P\) depends only on the state dimension \(n_x\) and importantly \emph{not} on the number of agents.
The same holds true for the LMI constraint~\eqref{eq:robust_stability_lmi} and the set \(\mathcal{P}_{n_x}\).
It is, however, necessary to verify \eqref{eq:robust_stability_lmi_modal} for all non-zero eigenvalues of \(L^0\), such that the number of constraints grows linearly with \(N\).

\subsection{Robust Performance}\label{sec:robust_performance}
The ideas outlined in the previous subsection can be applied in similar fashion to the \htwo{} analysis conditions in Theorem~\ref{thm:mjls_h2}.

\begin{theorem}[\htwo-Norm Bound]\label{thm:robust_h2}
    Given a decomposable MJLS~\(G\), a communication graph~\(\mathcal{G}^0\), and packet loss satisfying Assumptions~\ref{ass:markov} to \ref{ass:bounded_loss}, the MJLS is mean-square stable and \(\htwonorm{G} < \gamma\) if there exist \(Y \succ 0\), symmetric \(Z_i\) with \(\sum_{i = 1}^N \trace{Z_i} < \gamma^2\) and \(P_1, P_2 \in \mathcal{P}_{n_x + n_z}\) such that
    \begin{subequations}\label{eq:robust_h2_lmi}\begin{gather}
        A^{d\top} Y A^d - Y   + C^{d\top} C^d \prec 0 \label{eq:robust_h2_lmi_gramian_nominal} \\
        B^{d\top} Y B^d - Z_i + D^{d\top} D^d \prec 0 \label{eq:robust_h2_lmi_trace_nominal}
    \end{gather}\end{subequations}
    hold for all \(i\) such that \(\lambda_i = 0\) and \eqref{eq:robust_h2_lmi_gramian_modal}, \eqref{eq:robust_h2_lmi_trace_modal} hold for all non-zero \(\lambda_i\), where \(\lambda_i\) are the eigenvalues of \(L^0\)  and \(\mathcal{B}_j\) is the \(j\)th block-row of \(I_3 \otimes I_{n_x + n_z}\)
\end{theorem}
\begin{proof}
    \arxivTF{The steps that were outlined in the previous subsection can be applied analogously to Theorem~\ref{thm:mjls_h2}.
As the main difference, we have to evaluate the expectation not only of \(A_{\sigma_{k+1}}\) but in addition \(\expect{B_{\sigma_{k+1}} \given \sigma_k}\), \(\expect{C_{\sigma_{k+1}} \given \sigma_k}\), and \(\expect{D_{\sigma_{k+1}} \given \sigma_k}\).
Because these are all uncertain according to Assumption~\ref{ass:bounded_loss}, the uncertainty channel has dimension \(3 (n_x + n_z)\), and we introduce separate multipliers \(P_1, P_2 \in \mathcal{P}_{n_x + n_z}\) for \eqref{eq:mjls_h2_gramian_lmi} and \eqref{eq:mjls_h2_trace_lmi} since the FBSP has to be applied to both LMIs individually.
Furthermore, no structure has to be imposed on \(Z\) a priori, but a block-diagonal structure with \(Z_i\) on the diagonal is implied by the Schur complement.
The remaining steps are carried out analogously to the proof of Theorem~\ref{thm:robust_stability}.
}{Omitted for brevity, see \cite{ExtendedVersion}.}
\end{proof}

\begin{remark}
    The LMIs \eqref{eq:robust_stability_lmi_nominal} and \eqref{eq:robust_h2_lmi_gramian_nominal} are only feasible if \(A^d\) is Schur, i.e., all eigenvalues of \(A^d\) are strictly inside the unit circle.
    This condition is not satisfied for many relevant MAS control problems, e.g., consensus or formation control (cf. Section~\ref{sec:example}).
    In those cases, it is instrumental to verify the LMIs only for states in the subspace orthogonal to the vector of all ones, which is achieved by neglecting the 0 eigenvalue of \(L^0\) \cite{Fax2004, Hespe2022}.
    Thus, for MAS whose \(A^d\) matrix has a single eigenvalue at 1, it is sufficient to verify \eqref{eq:robust_stability_lmi_modal} for \(\lambda_2\) to \(\lambda_N\), ignoring only \(\lambda_1 = 0\), to guarantee mean-square stability or similarly satisfy \eqref{eq:robust_h2_lmi_gramian_modal} and \eqref{eq:robust_h2_lmi_trace_modal} on the same set of eigenvalues for upper bounding the \htwo-norm.
\end{remark}

\section{Application Example}\label{sec:example}
This section demonstrates the applicability and scalability of the proposed results on an example system.
The resulting LMI problems are solved in \name{Matlab} using \name{Yalmip} \cite{Loefberg2004} and source code for generating the figures is available at \cite{SourceCode}.

The agent model under consideration is a mass with friction, described in discrete-time as
\begin{align*}
    x_{k+1}^i &= \bbma
        1 & 1 \\
        0 & 0.1
    \ebma x_k^i + \bbma
        0 \\ 1
    \ebma u_k^i &
    y_k^i &= \bbma
        1 & 0
    \ebma x_k^i
\end{align*}
where \(x_k^i\), \(u_k^i\), and \(y_k^i\) are the state, input and output of agent~\(i\), respectively.
All agents implement the consensus protocol
\begin{equation*}
    u_k^i = w_k^i + \kappa \sum_{j \in \mathcal{N}_i^-} \theta_k^{ij} \big(y_k^j - y_k^i\big),
\end{equation*}
with gain \(\kappa > 0\).
Here, \(w_k^i\) is an external input disturbance acting on agent~\(i\).
By stacking up the states as \(x_k^\top = [x_k^{1\top}, \ldots, x_l^{N\top}]\), and similarly for the inputs and outputs, the MAS can be brought into the form of a decomposable MJLS, where we use \(z_k = L^0 y_k\) as performance output.
Note that \(z_k\) is a \emph{deterministic} performance channel because it is calculated using the nominal Laplacian \(L^0\).
This distinction is central to obtain meaningful results because a performance channel involving \(L_{\sigma_k}\) will vanish if \(\expect[\big]{\theta_k^{ij}}\) is small.

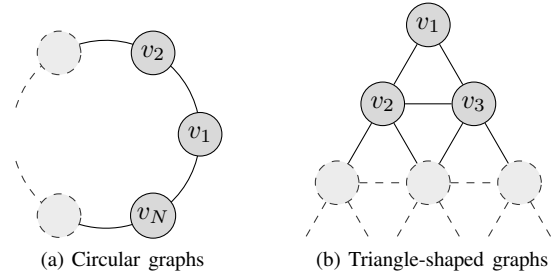
\begin{figure}
    \centering
    \subfloat[Circular graphs]{
        \centering
        \begin{tikzpicture}[scale=1.25]
    \draw (120:1) arc (120:-120:1);
    \draw[hidden] (120:1) arc (120:165:1);
    \draw[hidden] (240:1) arc (240:195:1);

	\draw (  0:1) node[agent](a1){$v_1$};
	\draw ( 60:1) node[agent](a2){$v_2$};
	\draw (120:1) node[hidden agent](a3){\phantom{$v_1$}};
	\draw (240:1) node[hidden agent](a4){\phantom{$v_1$}};
	\draw (300:1) node[agent](a5){\makebox[1em][c]{$v_N$}};
\end{tikzpicture}
        \label{fig:example_graphs_circular}
    }
    \hspace{6mm}
    \subfloat[Triangle-shaped graphs]{
        \centering
        \begin{tikzpicture}[scale=0.7]
    \pgfmathsetmacro{\ra}{cos(pi/6)}
    \pgfmathsetmacro{\rb}{2*\ra}

	\draw ( 30:\ra) node[agent](a3){$v_3$};
	\draw ( 90:\rb) node[agent](a1){$v_1$};
	\draw (150:\ra) node[agent](a2){$v_2$};
	\draw (210:\rb) node[hidden agent](a4){\phantom{$v_1$}};
	\draw (270:\ra) node[hidden agent](a5){\phantom{$v_1$}};
	\draw (330:\rb) node[hidden agent](a6){\phantom{$v_1$}};

	\draw (a1) -- (a2);
	\draw (a1) -- (a3);
	\draw (a2) -- (a3);
	\draw (a2) -- (a4);
	\draw (a2) -- (a5);
	\draw (a3) -- (a5);
	\draw (a3) -- (a6);
	\draw[hidden] (a4) -- (a5);
	\draw[hidden] (a5) -- (a6);

	\draw[hidden] (a4) -- ++($0.7*(a4)-0.7*(a2)$);
	\draw[hidden] (a5) -- ++($0.7*(a4)-0.7*(a2)$);
	\draw[hidden] (a6) -- ++($0.7*(a4)-0.7*(a2)$);
	\draw[hidden] (a4) -- ++($0.7*(a6)-0.7*(a3)$);
	\draw[hidden] (a5) -- ++($0.7*(a6)-0.7*(a3)$);
	\draw[hidden] (a6) -- ++($0.7*(a6)-0.7*(a3)$);
\end{tikzpicture}
        \label{fig:example_graphs_triangle}
    }
    \caption{Graph structures that are used in the example section.}
    \label{fig:example_graphs}
\end{figure}

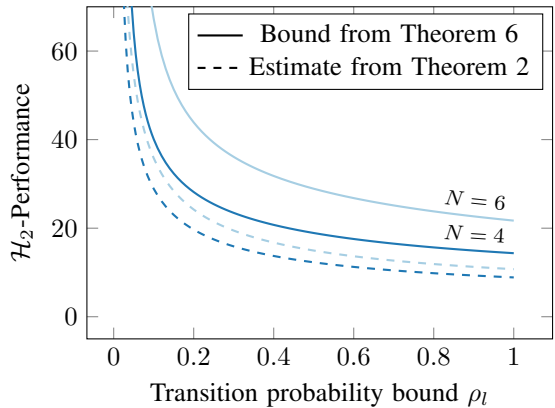
\begin{figure}
    \centering
    \begin{tikzpicture}
    \pgfplotstableread[col sep=comma]{figures/data/uncertainty_sweep.csv} \datatable

    \tikzset{
        path description/.style={
            black,
            font=\footnotesize,
            sloped,
            anchor=south east,
            pos=1
        }
    }

    \begin{axis}[
        width=0.9\columnwidth,
        height=6cm,
        xlabel=Transition probability bound $\rho_l$,
        ylabel=\htwo-Performance,
        ymin=-5, ymax=70
    ]
        \addlegendimage{color=black}
        \addlegendimage{color=black, dashed}

        \addplot table[x=p_swp, y=H2_dec_1] {\datatable} node[path description]{\(N = 4\)}; \label{fig:uncertainty_sweep_line_N4}
        \addplot table[x=p_swp, y=H2_dec_2] {\datatable} node[path description]{\(N = 6\)}; \label{fig:uncertainty_sweep_line_N6}
        \pgfplotsset{cycle list shift=-2>}
        \addplot+[dashed] table[x=p_swp, y=H2_enm_1] {\datatable};
        \addplot+[dashed] table[x=p_swp, y=H2_enm_2] {\datatable};

        \legend{Bound from Theorem~\ref{thm:robust_h2}, Estimate from Theorem~\ref{thm:mjls_h2}}
    \end{axis}

    \path (current axis.south west) -- ++(0, 0) rectangle (current axis.north east) -- ++(0, 1ex);
\end{tikzpicture}
    \caption{%
        Best upper bound on the \htwo-Performance that can be obtained from Theorem~\ref{thm:robust_h2} for transition probabilities in the interval \([\rho_l, 1]\) with \(N = 4\) (\ref{fig:uncertainty_sweep_line_N4}) and \(N = 6\) (\ref{fig:uncertainty_sweep_line_N6}).
        For comparison, an estimate obtained from Theorem~\ref{thm:mjls_h2} by evaluating all combinations of \(p^{ij} \in \set{\rho_l, 1}\) with the same \(X\) is included.
    }
    \label{fig:uncertainty_sweep}
\end{figure}

First, we demonstrate that the proposed analysis approach results in reasonably conservative upper bounds.
For circular graphs as shown in Figure~\ref{fig:example_graphs_circular}, we fix \(\rho_u = 1\) and let \(\rho_l\) vary between 0 and 1.
The resulting curves for four and six agents are plotted in Figure~\ref{fig:uncertainty_sweep}.
In addition, the figure contains an estimate of the \htwo-norm obtained by evaluating Theorem~\ref{thm:mjls_h2} for all combinations such that \(p^{ij} = q^{ij}\) is either \(\rho_l\) or 1 with the same \(X\).
Note that this estimate is exact for \(\rho_l = 1\).

As to be expected, the \htwo-norm bound is monotonically increasing for decreasing \(\rho_l\) because the uncertainty in the transmission probabilities is growing.
Furthermore, the gap between the bound from Theorem~\ref{thm:robust_h2} and the estimate from Theorem~\ref{thm:mjls_h2} grows with increased agent count.
This effect has already been observed in \cite{Hespe2022} and is caused by imposing the structure \(X = I_N \otimes Y\) for the matrix variables.

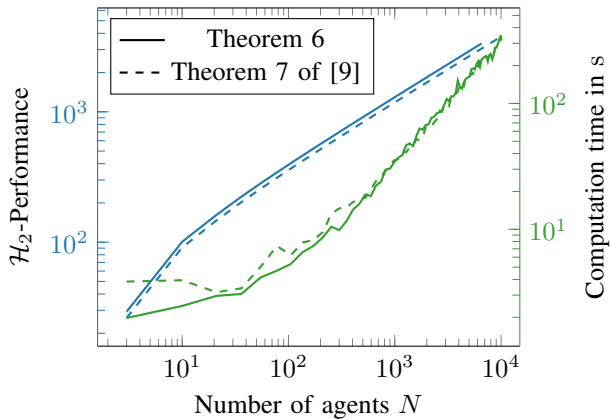
\begin{figure}
    \centering
    \begin{tikzpicture}
    \pgfplotstableread[col sep=comma]{figures/data/scaling.csv} \datatable

    \pgfplotsset{
        set layers,
        y axis style/.style={
            yticklabel style=#1,
            y axis line style=#1,
            ytick style=#1
        }
    }

    \begin{loglogaxis}[
        width=0.65\columnwidth,
        height=4.5cm,
        scale only axis,
        axis y line*=left,
        y axis style=Paired-B,
        xlabel=Number of agents $N$,
        ylabel=\htwo-Performance,
        xmin=1.6, xmax=15000,
        legend pos=north west
    ]
        \addlegendimage{color=black}
        \addlegendimage{color=black, dashed}

        \addplot[Paired-B] table[x=N, y=H2_rbst] {\datatable};
        \addplot[Paired-B, dashed] table[x=N, y=H2_sing] {\datatable};

        \legend{Theorem~\ref{thm:robust_h2}, Theorem~7 of \cite{Hespe2022}}
    \end{loglogaxis}

    \begin{loglogaxis}[
        width=0.65\columnwidth,
        height=4.5cm,
        scale only axis,
        axis y line*=right,
        axis x line=none,
        y axis style=Paired-D,
        ylabel=Computation time in \si{\second},
        xmin=1.6, xmax=15000
    ]
        \addplot[Paired-D] table[x=N, y=time_rbst] {\datatable};
        \addplot[Paired-D, dashed] table[x=N, y=time_sing] {\datatable};
    \end{loglogaxis}

    \path (current axis.south west) -- ++(0, 0) rectangle (current axis.north east) -- ++(0, 1ex);
\end{tikzpicture}
    \caption{%
        \htwo-Performance as obtained from Theorem~\ref{thm:robust_h2} for different agent counts with transition probabilities in \([0.4, 0.6]\).
        The second axis shows the computation time required to solve the optimization problem, averaged over ten runs.
        \htwo-Performance and computation time as obtained from Theorem~7 in \cite{Hespe2022} for \(p = 0.5\) are shown for reference.
    }
    \label{fig:scalability}
\end{figure}

The second scenario studies how the approach fares with increasing size of the MAS.
Starting with \(N = 3\), Theorem~\ref{thm:robust_h2} is applied to MAS with triangle-shaped communication structure as shown in Figure~\ref{fig:example_graphs_triangle} and \(p^{ij} = q^{ij} \in [0.4, 0.6]\).
For comparison, we apply Theorem~7 of \cite{Hespe2022} with fixed and \emph{homogeneous} \(p = 0.5\).
The best obtainable upper bounds and the required computation times are shown in Figure~\ref{fig:scalability}.

The gap between the two \htwo-performance curves in the figure stems from treating the transmission probabilities as uncertain in Theorem~\ref{thm:robust_h2} and is independent of network size on the shown logarithmic scale.
Importantly, the computational cost of both approaches scales well with the number of agents, such that even with 10,000 agents, calculating a robust upper bound on the \htwo-norm takes only a few minutes.

\section{Conclusions}\label{sec:conclusions}
This paper presents a scalable approach to obtain a sufficient mean-square stability condition and upper bounds on the \htwo-norm for MAS with stochastic packet loss.
Different from previous results, the conditions remain valid if the packet loss is correlated in time or has heterogeneous probabilities across the MAS.
An important feature of the approach is that its complexity scales linearly with the number of agents and therefore application to very large MASs is tractable.

\bibliographystyle{IEEETran}
\bibliography{arxiv}

\end{document}